\documentclass[a4paper]{article}
\usepackage{interspeech2010,amssymb,amsmath,epsfig}
\usepackage{graphicx}
\usepackage{multirow}

\ninept
\def\reg{{\rm\ooalign{\hfil
     \raise.07ex\hbox{\scriptsize R}\hfil\crcr\mathhexbox20D}}}

\title{Analysis and Synthesis of Hypo and Hyperarticulated Speech}

\makeatletter
\def\name#1{\gdef\@name{#1\\}}
\makeatother
\name{{\em Benjamin Picart, Thomas Drugman, Thierry Dutoit}}

\address{TCTS Lab, Facult\'e Polytechnique (FPMs), University of Mons (UMons), Belgium\\
{\small \tt \{benjamin.picart,thomas.drugman,thierry.dutoit\}@umons.ac.be}}

\begin{document}
\maketitle

\begin{abstract}
This paper focuses on the analysis and synthesis of hypo and hyperarticulated speech in the framework of HMM-based speech synthesis. First of all, a new French database matching our needs was created, which contains three identical sets, pronounced with three different degrees of articulation: neutral, hypo and hyperarticulated speech. On that basis, acoustic and phonetic analyses were performed. It is shown that the degrees of articulation significantly influence, on one hand, both vocal tract and glottal characteristics, and on the other hand, speech rate, phone durations, phone variations and the presence of glottal stops. Finally, neutral, hypo and hyperarticulated speech are synthesized using HMM-based speech synthesis and both objective and subjective tests aiming at assessing the generated speech quality are performed. These tests show that synthesized hypoarticulated speech seems to be less naturally rendered than neutral and hyperarticulated speech.

\end{abstract}

\section{Introduction}\label{sec:intro}

In this paper, we focus on the study of different speech styles, based on the degree of articulation: neutral speech, hypoarticulated (or casual) and hyperarticulated speech (or clear speech). It is worth noting that these three modes of expressivity are neutral on the emotional point of view, but can vary amongst speakers, as reported in \cite{Beller_thesis}. The influence of emotion on the articulation degree has been studied in \cite{Beller_expressivite}, \cite{Beller_expressivite2} and is out of the scope of this work.


The ``H and H" theory \cite{Lindblom} proposes two degrees of articulation of speech: hyperarticulated speech, for which speech clarity tends to be maximized, and hypoarticulated speech, where the speech signal is produced with minimal efforts. Therefore the degree of articulation provides information on the motivation/personality of the speaker vs the listeners \cite{Beller_thesis}. Speakers can adopt a speaking style that allows them to be understood more easily in difficult communication situations.
The degree of articulation is influenced by the phonetic context, the speech rate and the spectral dynamics (vocal tract rate of change). The common measure of the degree of articulation consists in defining formant targets for each phone, taking coarticulation into account, and studying the differences between the real observations and the targets versus the speech rate. Because defining formant targets is not an easy task, Beller proposed in \cite{Beller_thesis} a statistical measure of the degree of articulation by studying the joint evolution of the vocalic triangle area and the speech rate.


The goal of this study is to have a better understanding of the specific characteristics (acoustic and phonetic) governing hypo and hyperarticulated speech and to apply it to HMM synthesis. In order to achieve this goal, the paper is divided into two main parts: the analysis (Section \ref{sec:Analysis}) and synthesis (Section \ref{sec:Synthesis}) of hypo and hyperarticulated speech.

In the first part, the acoustic (Section \ref{ssec:Acoustic}) and phonetic (Section \ref{ssec:Phonetic}) modifications are studied as a function of the degree of articulation. The acoustic analysis highlights evidence of both vocal tract and glottal characteristics changes, while the phonetic analysis focuses on showing evidence of glottal stops presence, phone variations, phone durations and speech rate changes. In the second part, the integration within a HMM-based speech synthesizer in order to generate the two degrees of articulation is discussed (Section \ref{ssec:HTS_Modif}). Both an objective and subjective evaluation are carried out with the aim of assessing how the synthetic speech quality is affected for both degrees of articulation. Finally Section \ref{sec:conclu} concludes the paper and some of our future works are given in Section \ref{sec:FutureWork}.

\section{Creation of a Database with various Degrees of Articulation}\label{sec:dba}

For the purpose of our research, a new French database was recorded by a professional male speaker, aged 25 and native French (Belgium) speaking. The database contains three separate sets, each set corresponding to one degree of articulation (neutral, hypo and hyperarticulated). For each set, the speaker was asked to pronounce the same 1359 phonetically balanced sentences, as neutral as possible from the emotional point of view. A headset was provided to the speaker for both hypo and hyperarticulated recordings, in order to induce him to speak naturally while modifying his articulation degree.

While recording hyperarticulated speech, the speaker was listening to a version of his voice modified by a ``Cathedral" effect. This effect produces a lot of reverberations (as in a real cathedral), forcing the speaker to talk slower and as clearly as possible (more efforts to produce speech). On the other hand, while recording hypoarticulated speech, the speaker was listening to an amplified version of his own voice. This effect produces the impression of talking very close to someone in a narrow environment, allowing the speaker to talk faster and less clearly (less efforts to produce speech). Proceeding that way allows us to create a ``standard recording protocol" to obtain repeatable conditions if required in the future.


\section{Analysis of Hypo and Hyperarticulated Speech}\label{sec:Analysis}

\subsection{Acoustic Analysis}\label{ssec:Acoustic}

Acoustic modifications in expressive speech have been extensively studied in the literature \cite{Klatt}, \cite{Childers}, \cite{Keller}. In the frame of this study, one can expect important changes related to the vocal tract function. Indeed, during the production of hypo and hyperarticulated speech, the articulatory strategy adopted by the speaker may dramatically vary. Although it is still not clear whether these modifications consist of a reorganization of the articulatory movements, or of a reduction/amplification of the normal ones, speakers generally tend to consistently change their way of articulating. According to the ``H and H" theory \cite{Lindblom}, speakers minimize their articulatory trajectories in hypoarticulated speech, resulting in a low intelligibility, while an opposite strategy is adopted in hyperarticulated speech. As a consequence, the vocal tract configurations may be strongly affected. The resulting changes are studied in Section \ref{sssec:VT}.

In addition, the produced voice quality is also altered. Since voice quality variations are mainly considered to be controlled by the glottal source \cite{Keller}, Section \ref{sssec:Glottal} focuses on the modifications of glottal characteristics with regard to the degree of articulation.

\subsubsection{Vocal Tract-based Modifications}\label{sssec:VT}
\begin{figure}[!ht]
  \centering
  \includegraphics[width=0.45\textwidth]{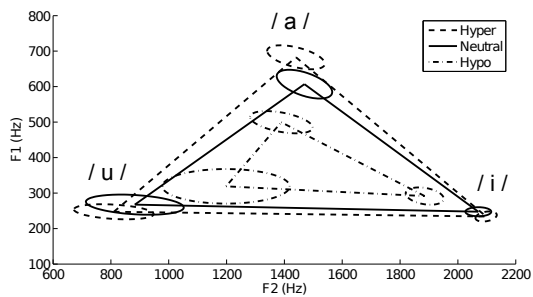}
  \caption{Vocalic triangle, for the three degrees of articulation, estimated on the \textbf{original} recordings. Dispersion ellipses are also indicated.}
  \label{fig:triangle_vocalique_original}
\end{figure}

In order to study the variations of the vocal tract resonances, the evolution of the vocalic triangle \cite{Beller_thesis} with the degree of articulation was analyzed. This triangle consists of the three vowels $/a/$, $/i/$ and $/u/$ represented in the space of the two first formant frequencies $F1$ and $F2$ (here estimated via Wavesurfer \cite{Wavesurfer}). For the three degrees of articulation, the vocalic triangle is displayed in Figure \ref{fig:triangle_vocalique_original} for the original sentences. For information, ellipses of dispersion are also indicated on these plots. The first main conclusion is the significant reduction of the vocalic space as speech becomes less articulated. Indeed, as the articulatory trajectories are less marked, the resulting acoustic targets are less separated in the vocalic space. This may partially explain the lowest intelligibility in hypoarticulated speech. On the contrary, the enhanced acoustic contrast is the result of the efforts of the speaker under hyperarticulation. These changes of vocalic space are summarized in Table \ref{tab:TriangleArea}, which presents the area defined by the average vocalic triangles.



\begin{table}[!ht]
\centering
\begin{tabular}{ c || c | c | c }
\hline
Dataset & Hyper & Neutral & Hypo \\
\hline
\hline
Original & 0.285 & 0.208 & 0.065 \\
\hline
\end{tabular}
\caption{Vocalic space (in $kHz^2$) for the three degrees of articulation for the original sentences.}
\label{tab:TriangleArea}
\end{table}

Inspecting the ellipses, it is observed that dispersion can be high for the vowel $/u/$, while data is relatively well concentrated for $/a/$ and $/i/$.

\subsubsection{Glottal-based Modifications}\label{sssec:Glottal}
\begin{figure}[!ht]
  \centering
  \includegraphics[width=0.45\textwidth]{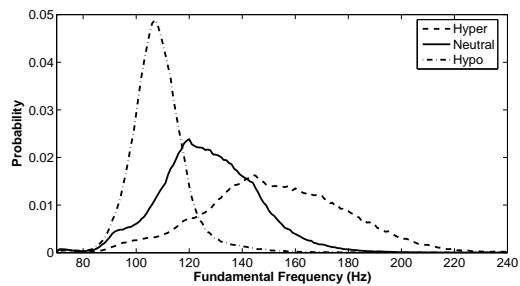}
  \caption{Pitch histograms for the three degrees of articulation.}
  \label{fig:PitchHisto}
\end{figure}

\begin{figure}[!ht]
  \centering
  \includegraphics[width=0.45\textwidth]{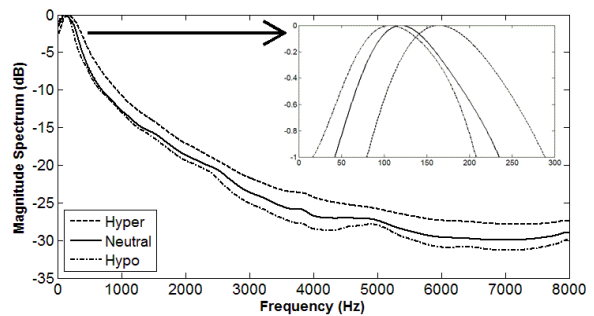}
  \caption{Averaged magnitude spectrum of the glottal source for the three degrees of articulation.}
  \label{fig:MaxSpec}
\end{figure}

As the most important perceptual glottal feature, pitch histograms are displayed in Figure \ref{fig:PitchHisto}. It is clearly noted that the more speech is articulated, the higher the fundamental frequency. Besides these prosodic modifications, we investigate how characteristics of the glottal flow are affected. In a first part, the glottal source is estimated by the Complex Cepstrum-based Decomposition algorithm (CCD, \cite{CCD}). This method relies on the mixed-phase model of speech \cite{MixedPhase}. According to this model, speech is composed of both minimum-phase and maximum-phase components, where the latter contribution is only due to the glottal flow. By isolating the maximum-phase component of speech, the CCD method has shown its ability to efficiently estimate the glottal source. Using this technique, Figure \ref{fig:MaxSpec} shows the averaged magnitude spectrum of the glottal source for the three degrees of articulation. First of all, a strong similarity of these spectra with models of the glottal source (such as the LF model \cite{LF}) can be noticed. Secondly it turns out that a high degree of articulation is reflected by a glottal flow containing a greater amount of high frequencies. Finally, it is also observed that the glottal formant frequency increases with the degree of articulation (see the zoom in the top right corner of Figure \ref{fig:MaxSpec}). In other words, the time response of the glottis open phase turns to be faster in hyperarticulated speech.

\begin{figure}[!ht]
  \centering
  \includegraphics[width=0.45\textwidth]{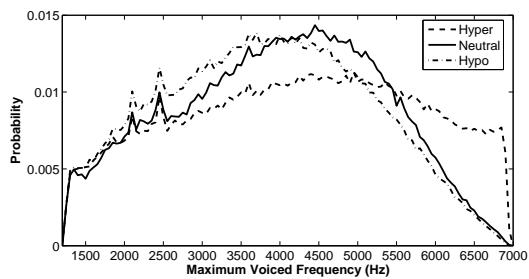}
  \caption{Histograms of the maximum voiced frequency for the three degrees of articulation.}
  \label{fig:Fm}
\end{figure}

In a second part, the maximum voiced frequency is analyzed. In some approaches, such as the Harmonic plus Noise Model (HNM, \cite{HNM}) or the Deterministic plus Stochastic Model of residual signal (DSM, \cite{DSM}) which will be used for synthesis in Section \ref{sec:Synthesis}, the speech signal is considered to be modeled by a non-periodic component beyond a given frequency. This maximum voiced frequency ($F_m$) demarcates the boundary between two distinct spectral bands, where respectively an harmonic and a stochastic modeling (related to the turbulences of the glottal airflow) are supposed to hold. In this paper, $F_m$ was estimated using the algorithm described in \cite{HNM}. The corresponding histograms are illustrated in Figure \ref{fig:Fm} for the three degrees of articulation. It can be noticed from this figure that the more speech is articulated, the higher the $F_m$, the stronger the harmonicity, and consequently the weaker the presence of noise in speech. Note that the average values of $F_m$ are respectively of 4215 Hz, 3950 Hz (confirming our choice of 4 kHz in \cite{DSM}) and 3810 Hz for the three degrees of articulation.

\subsection{Phonetic Analysis}\label{ssec:Phonetic}
Phonetic modifications in hypo and hyperarticulated speech are also very important characteristics to investigate. In the next paragraphs, glottal stops (Section \ref{sssec:glottalstop}), phone variations (Section \ref{sssec:phonevar}), phone durations (Section \ref{sssec:phonedur}) and speech rates (Section \ref{sssec:speechrate}) are analyzed. In order to obtain reliable results, the entire database for each degree of articulation is used in this section. Moreover, the 36 standard French phones are considered (\cite{phonetic} from which \emph{/\^{a}/} and \emph{/ng/} are not used because they can be made from other phonemes, and \emph{/\_ /} is added). Note that results can vary from one speaker to another as pointed out in \cite{Beller_thesis}. Eventually, the database was segmented using HMM forced alignment \cite{Malfrere2003}.

\subsubsection{Glottal Stops}\label{sssec:glottalstop}
\begin{figure}[!ht]
  \centering
  \includegraphics[width=0.45\textwidth]{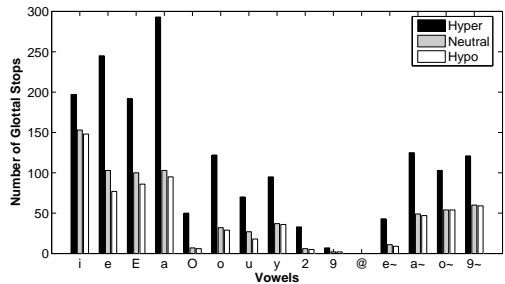}
  \caption{Number of glottal stops for each phone (vowel) and each degree of articulation.}
  \label{fig:glottal_stop}
\end{figure}

According to \cite{BBC}, a glottal stop is a cough-like explosive sound released just after the silence produced by the complete glottal closure.
In French, such a phenomenon happens when the glottis closes completely before a vowel. A method for detecting glottal stops in continuous speech was proposed in \cite{Yegnanarayana}. However, this technique was not used here. Instead we detected glottal stops manually. Figure \ref{fig:glottal_stop} shows, for each vowel, the number of glottal stops for each degree of articulation. It turns out from this figure that the number of glottal stops is much higher (almost always double) in hyperarticulated speech than in neutral and hypoarticulated speech (between which no sensible modification is noticed).


\subsubsection{Phone Variations}\label{sssec:phonevar}

Phone variations refer to phonetic insertions, deletions and substitutions that the speaker makes during hypo and hyperarticulation, compared to the neutral speech. This study has been performed at the phone level, considering the phone position in the word, and at the phone group level (groups of phones that were inserted, deleted or substituted).

For the sake of conciseness, only the most relevant differences will be given in this section. Table \ref{tab:phonevaria} shows, for each phone, the total proportion of phone deletions in hypoarticulated speech and phone insertions in hyperarticulated speech (first line). The position of these deleted/inserted phones inside the words are also shown: at the beginning (second line), in the middle (third line) and at the end (fourth line). Note that since there is no significant deletion process in hyperarticulation, no significant insertion process in hypoarticulation and no significant substitution process in both cases, they do not appear in Table \ref{tab:phonevaria}.

\begin{table*}[!ht]
\centering
\begin{tabular}{c | c || c | c | c | c | c | c | c | c | c | c | c }
\hline
\multicolumn{2}{ c || }{Phone} & $/j/$ & $/H/$ & $/t/$ & $/k/$ & $/z/$ & $/Z/$ & $/l/$ & $/R/$ & $/E/$ & $/@/$ & $/\_/$ \\
\hline
\hline
\multirow{4}{*}{} & Tot & 1.5 & 1.7 & 1.9 & 1.6 & \textbf{3.1} & \textbf{5.1} & \textbf{2.2} & \textbf{3.4} & 1.5 & \textbf{29.7} & \textbf{14.2} \\
\cline{2-13}
Deletions & Beg & 0.14 & 0.57 & 0.14 & 0.38 & 0.0 & \textbf{4.95} & 0.26 & 0.03 & 0.89 & 11.49 & \textbf{14.2} \\
\cline{2-13}
(Hypoarticulation) & Mid & 0.53 & 1.13 & 0.52 & 0.45 & 0.94 & 0.15 & 0.44 & 1.62 & 0.47 & 2.85 & 0.0 \\
\cline{2-13}
& End & 0.82 & 0.0 & 1.24 & 0.77 & \textbf{2.16} & 0.0 & \textbf{1.50} & \textbf{1.75} & 0.14 & \textbf{15.39} & 0.0 \\
\hline
\hline
\multirow{4}{*}{} & Tot & 0.3 & 0.0 & 1.1 & 0.2 & 4.0 & 0.6 & 0.1 & 0.2 & 0.2 & \textbf{40.0} & \textbf{26.5} \\
\cline{2-13}
Insertions & Beg & 0.0 & 0.0 & 0.10 & 0.0 & 0.0 & 0.0 & 0.025 & 0.0 & 0.05 & 0.60 & \textbf{26.5} \\
\cline{2-13}
(Hyperarticulation) & Mid & 0.15 & 0.0 & 0.25 & 0.07 & 0.41 & 0.15 & 0.025 & 0.04 & 0.1 & 1.68 & 0.0 \\
\cline{2-13}
& End & 0.15 & 0.0 & 0.75 & 0.13 & 3.59 & 0.45 & 0.05 & 0.16 & 0.05 & \textbf{37.72} & 0.0 \\
\hline

\end{tabular}
\caption{Total percentage (first line) of deleted and inserted phones in hypo and hyperarticulated speech respectively, and their repartition inside the words: beginning (second line), middle (third line), end (fourth line).}
\label{tab:phonevaria}
\end{table*}


In hyperarticulated speech, the only important insertions are breaks $/\_ /$ and Schwa $/@/$ (mostly at the end of the words). In hypoarticulated speech, breaks and Schwa (mostly at the end of the words) are often deleted, as $/R/$, $/l/$, $/Z/$ and $/z/$. Schwa, also called ``mute e" or ``unstable e", is very important in French. It is the only vowel that can or cannot be pronounced (all other vowels should be clearly pronounced), and several authors have focused on Schwa insertions and deletions in French. The analysis performed at the phone group level is still under development but we observed frequent phone group deletions in hypoarticulated speech (e.g. $/R@/$, $/l@/$ at the end of the words, \emph{/je suis/} (which means \emph{/I am/}) becoming \emph{/j'suis/} or even \emph{/chui/}, ...) and no significant group insertions in hyperarticulated speech. In both cases, no significant phone groups substitutions were observed.

\subsubsection{Phone Durations}\label{sssec:phonedur}
Intuitively, it is expected that the degree of articulation has an effect on phone durations, as well as on the speech rate (Section \ref{sssec:speechrate}). Some studies are confirming that thought. In the approach exposed in \cite{Jurafsky}, it is found evidence for the Probabilistic Reduction Hypothesis: word forms are reduced when they have a higher probability, and this should be interpreted as evidence that probabilistic relations between words are represented in the mind of the speaker. Similarly, \cite{Bradlow} examines how that probability (lexical frequency and previous occurrence), speaking style, and prosody affect word duration, and how these factors interact.

\begin{figure}[!ht]
  \centering
  \includegraphics[width=0.45\textwidth]{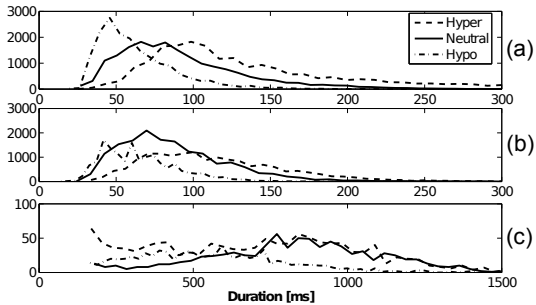}
  \caption{Phone durations histograms. (a) front, central, back \& nasal vowels. (b) plosive \& fricative consonants. (c) breaks.}
  \label{fig:phone_durations1}
\end{figure}

\begin{figure}[!ht]
  \centering
  \includegraphics[width=0.45\textwidth]{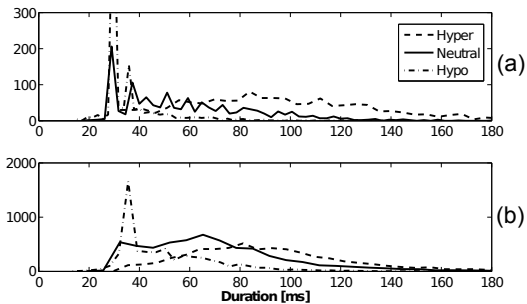}
  \caption{Phone durations histograms. (a) semi-vowels. (b) trill consonants.}
  \label{fig:phone_durations2}
\end{figure}

In this work, we have investigated the phone duration variations between neutral, hypoarticulated and hyperarticulated speech. Vowels and consonants were grouped according to broad phonetic classes \cite{phonetic}. Figure \ref{fig:phone_durations1} shows the histograms of (a) front, central, back and nasal vowels, (b) plosive and fricative consonants, and (c) breaks. Figure \ref{fig:phone_durations2} shows the histograms of (a) semi-vowels and (b) trill consonants. As expected, one can see that, generally, phone durations are shorter in hypoarticulation and longer in hyperarticulation. Breaks are shorter (and more rare) in hypoarticulation, but are as long as the ones in neutral speech and more present in hyperarticulation. An interesting characteristic of hypoarticulated speech is the concentration (high peaks) of semi-vowels and trill consonants in the short durations.

\subsubsection{Speech Rate}\label{sssec:speechrate}
Speaking rate has been found to be related to many factors \cite{Yuan}. It is often defined as the average number of syllables uttered per second (pauses excluded) in a whole sentence \cite{Beller_speechrates}, \cite{Roekhaut}. Based on that definition, Table \ref{tab:speechrate} compares the three speaking styles. 

\begin{table}[!ht]
\centering
\begin{tabular}{ c || c | c | c }
\hline
Results & Hyper & Neutral & Hypo \\
\hline
\hline
Total speech time [s] & 6076 & 4335 & 2926 \\
\hline
Total syllable time [s] & 5219 & 3618 & 2486 \\
\hline
Total pausing time [s] & 857 & 717 & 440 \\
\hline
Total number of syllables & 19736 & 18425 & 17373 \\
\hline
Total number of breaks & 1213 & 846 & 783 \\
\hline
Speech rate [syllable/s] & 3.8 & 5.1 & 7.0 \\
\hline
Pausing time [\%] & 14.1 & 16.5 & 15.1 \\
\hline
\end{tabular}
\caption{Results for hypo, neutral \& hyperarticulated speech.}
\label{tab:speechrate} 
\end{table}


As expected, hyperarticulated speech is characterized by a lower speech rate, a higher number of breaks (thus a longer pausing time), more syllables (final Schwa insertions), resulting in an increase of the total speech time. 

On the other side, hypoarticulated speech is characterized by a higher speech rate, a lower number of breaks (thus a shorter pausing time), less syllables (final Schwa and other phone groups deletions), resulting in a decrease of the total speech time. An interesting property can be noted: because of the increase (decrease) in the total pausing time and the total speech time in hyper (hypo) articulated speech, the pausing time (thus the speaking time) expressed in percents of the total speech time is almost independent of the speech style.

\section{Synthesis of Hypo and Hyperarticulated Speech}\label{sec:Synthesis}
Synthesis of the articulation degree in concatenative speech synthesis has been performed in \cite{Wouters}, by modifying the spectral shape of acoustic units according to a predictive model of the acoustic-prosodic variations related to the articulation degree. In this paper, we report our first attempts in synthesizing the two degrees of articulation of speech using HMM-based speech synthesis (via HTS \cite{HTS}).

\subsection{Integration within HMM-based Speech Synthesis}\label{ssec:HTS_Modif}
%

For each degree of articulation, a HMM-based speech synthesizer \cite{HTS_ref} was built, relying for the implementation on the HTS toolkit (version 2.1) publicly available in \cite{HTS}. In each case, 1220 sentences sampled at 16kHz were used for the training, leaving around 10\% of the database for synthesis. For the filter, we extracted the traditional Mel Generalized Cepstral coefficients (with frequency warping factor = 0.42, gamma = 0 and order of MGC analysis = 24). For the excitation, we used the Deterministic plus Stochastic Model (DSM) of the residual signal proposed in \cite{DSM}, since it was shown to significantly improve the naturalness of the delivered speech. More precisely, both deterministic and stochastic components of the DSM model were estimated from the training dataset for each degree of articulation. The spectral boundary between these two components was chosen as the averaged value of the maximum voiced frequency described in Section \ref{sssec:Glottal}.

The objective of this preliminary work was to assess the quality of the synthesized speech based only on phonetic transcription modifications. Therefore, hypo and hyperarticulated speech were obtained by manually modifying the phonetic transcriptions at the input of the synthesizer, according to Section \ref{sssec:phonevar} (our future natural language processor should do it automatically).
In the following evaluations, original pitch and phone durations were imposed at the input of the synthesizers.


\subsection{Acoustic Analysis}\label{ssec:Acoustic_Synth}
\begin{figure}[!ht]
  \centering
  \includegraphics[width=0.45\textwidth]{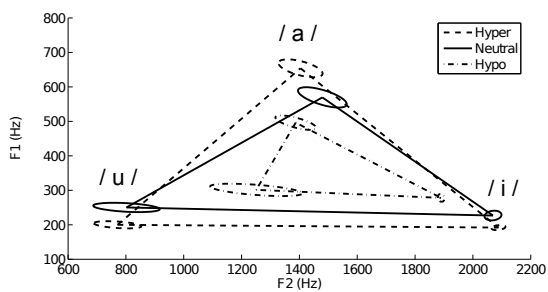}
  \caption{Vocalic triangle, for the three degrees of articulation, estimated on the sentences \textbf{generated} by the HMM-based synthesizer. Dispersion ellipses are also indicated.}
  \label{fig:triangle_vocalique_synthese}
\end{figure}

The same acoustic analysis as in Section \ref{sssec:VT} was performed on the sentences generated by the HMM-based synthesizer. Results are summarized in Figure \ref{fig:triangle_vocalique_synthese} and in Table \ref{tab:TriangleArea_Synth}. Note the good agreement between vocalic spaces in original (see Section \ref{sssec:VT}) and synthesized sentences.

\begin{table}[!ht]
\centering
\begin{tabular}{ c || c | c | c }
\hline
Dataset & Hyper & Neutral & Hypo \\
\hline
\hline
Synthesis & 0.302 & 0.210 & 0.064 \\
\hline
\end{tabular}
\caption{Vocalic space (in $kHz^2$) for the three degrees of articulation for the synthesized sentences.}
\label{tab:TriangleArea_Synth}
\end{table}

The same conclusions as in Section \ref{sssec:VT} hold for the synthetic examples. In other words, the essential vocalic characteristics are preserved despite the HMM-based modeling and generation process. It can be however noticed that the dispersion of the formant frequencies is lower after generation, especially for $F1$. This is mainly due to an over-smoothing of the generated spectra (albeit the Global Variance method \cite{GV} was used).

\subsection{Objective Evaluation}\label{ssec:Object_Eval}
The goal of the objective evaluation is to assess whether HTS is capable of producing natural hypo and hyperarticulated speech and to which extent. The distance measure considered here is the mel-cepstral distortion between the target and the estimated mel-cepstra coefficients, expressed as: 

\begin{equation}
Mel-CD = \frac{10}{ln(10)}\sqrt{2\sum_{d=1}^{24} (mc_{d}^{(t)} - mc_{d}^{(e)})^{2}}
\end{equation}

This mel-cepstral distortion is computed for all the vowels of the database. Table \ref{tab:objecteval} shows the mean with its 95\% confidence interval for each degree of articulation. This objective evaluation shows that the mel-cepstral distortion increases from hyper to hypoarticulated speech.


\begin{table}[!ht]
\centering
\begin{tabular}{c || c | c | c }
\hline
Results & Hyper & Neutral & Hypo \\
\hline
\hline
Mean $\pm$ CI & 5.9 $\pm$ 0.1 & 6.3 $\pm$ 0.2 & 6.9 $\pm$ 0.1 \\
\hline
\end{tabular}
\caption{Objective evaluation results (in [dB]): mean score with its 95\% confidence interval (CI) for each degree of articulation.}
\label{tab:objecteval}
\end{table}


\subsection{Subjective Evaluation}\label{ssec:Subject_Eval}

In order to confirm the objective evaluation conclusion, we performed a subjective evaluation. For this evaluation, the listener was asked to compare three sentences: A, the original; B, the sentence vocoded by DSM; C, the sentence synthesized by HTS using DSM as vocoder. He was asked to score, on a 9-point scale, the overall speech quality of B in comparison with A and C. B was allowed to vary from 0 (= same quality as A) to 9 (= same quality as C). Therefore this score should be interpreted in terms of a ``distance" between B and A and C: the lower the score, the more B ``sounds like" A and thus the better the quality, and conversely.

The test consists of 15 triplets (5 sentences per degree of articulation), giving a total of 45 sentences. Before starting the test, the listener was provided with some reference sentences covering most of the variations to help him familiarize with the scale. During the test, he was allowed to listen to the triplet of sentences as many times as he wanted, in the order he preferred (he was advised to listen to A and C before listening to B, in order to know the boundaries). However he was not allowed to come back to previous sentences after validating his decision.

\begin{figure}[!ht]
\centering
\includegraphics[width=0.45\textwidth]{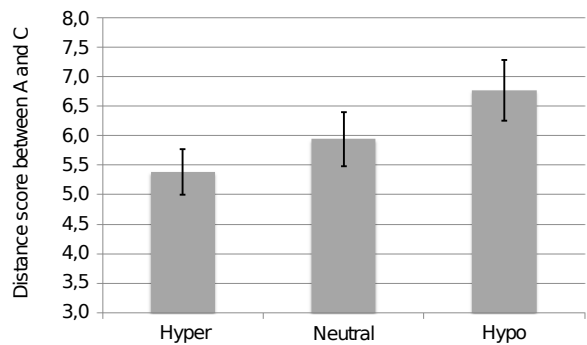}
\caption{Subjective evaluation results: overall speech quality of the HMM-based speech synthesizer (mean score with its 95\% confidence interval for each degree of articulation).}
\label{fig:subj_eval}
\end{figure}

The hypothesis made in this subjective evaluation is that the distance between A and B is constant, whatever the degree of articulation is. This hypothesis has been verified by informal listening. By proceeding this way, speech quality of C vs A can be assessed indirectly. 26 people, mainly naive listeners, participated to this evaluation. The mean score, corresponding to the ``distance" between A and C, together with its 95\% confidence interval for each articulation degree, on the 9-point scale, is shown in Figure \ref{fig:subj_eval}. The lower the score, the more C ``sounds like" A and thus the better the quality, and conversely. One can see that hypoarticulated speech is the worst, followed by neutral and hyperarticulated speech, therefore confirming the objective evaluation result.

\section{Conclusion}\label{sec:conclu}
This work is a first approach towards HMM-based hyper and hypoarticulated speech synthesis. A new French database matching our needs was created: three identical sets, pronounced with three different degrees of articulation (neutral, hypo and hyperarticulated speech).

In a first step, acoustic and phonetic analyses were performed on these databases, and the influence of the articulation degree on various factors was studied. It was shown that hyperarticulated speech is characterized by a larger vocalic space (more efforts to produce speech, with maximum clarity), higher fundamental frequency, a glottal flow containing a greater amount of high frequencies and an increased glottal formant frequency, the presence of a higher number of glottal stops, breaks and syllables, significant phone variations (especially insertions), longer phone durations and lower speech rate. The opposite tendency was observed in hypoarticulated speech, except that the number of glottal stops was equivalent to the one in neutral speech and the significant phone variations were deletions.

In a second step, synthesizing hypo and hyperarticulated speech was performed using HTS, based on modifications of the phonetic transcriptions at the input of the synthesizer, and of the characteristics of the excitation modeling. Objective and subjective evaluations were proposed in order to assess the quality of the synthesized speech. These tests show that the worst speech quality was obtained for hypoarticulated speech.

Audio examples for each degree of articulation are available online via http://tcts.fpms.ac.be/{\raise.17ex\hbox{$\scriptstyle\sim$}}picart/HypoAndHyperarticu- latedSpeech\_Demo.html.


\section{Discussion and Future Works}\label{sec:FutureWork}
The ultimate goal of our research is to be able to synthesize hypo and hyperarticulation, directly from an existing neutral voice (using voice conversion), without requiring recordings of new hypo and hyperarticulated databases (as done in this work). Right now, as the objective and subjective evaluations showed, the HMM-based speech synthesizers are not able to synthesize hypo and hyperarticulated speech with the same quality, even using the real hypo and hyperarticulated databases. It is therefore worth focusing on improving the current synthesis method before starting the next step: speaking style conversion. We will first investigate the simple methods for improving speaker-similarity in HMM-based speech synthesis proposed by \cite{Yamagishi}.



\section{Acknowledgments}\label{sec:Acknowledgments}

Benjamin Picart is supported by the ``Fonds pour la formation \`{a} la Recherche dans l'Industrie et dans l'Agriculture'' (FRIA). Thomas Drugman is supported by the ``Fonds National de la Recherche Scientifique'' (FNRS). Authors would like to thank Y. Stylianou for providing the algorithm for extracting $F_m$, and also Acapela Group SA for providing useful advices on database recordings and helping us in segmenting the database.

\eightpt
\bibliographystyle{IEEEtran}

\end{document}